\documentclass[
  aps,
  prl,
  twocolumn,
  superscriptaddress,
  floatfid,
  nofootinbib
]{revtex4-2}

\usepackage{amsmath,amssymb}      
\usepackage{graphicx}             
\usepackage[colorlinks=true,       
            linkcolor=blue,
            citecolor=blue,
            urlcolor=blue]{hyperref}
\usepackage{booktabs}   
\usepackage[caption=false]{subfig}
                     
\graphicspath{{paper_figs/}}

\usepackage{float}               
\begin{document}

\title{Benchmarking a Tunable Quantum Neural Network \\ on Trapped-Ion and Superconducting  Hardware}

\author{Djamil Lakhdar-Hamina}
\affiliation{Joint Quantum Institute and Department of Physics, University of Maryland, College Park, Maryland 20742, USA}  

\author{Xingxin Liu}
\affiliation{Joint Quantum Institute and Department of Physics, University of Maryland, College Park, Maryland 20742, USA} 

\author{Richard Barney}
\affiliation{Joint Quantum Institute and Department of Physics, University of Maryland, College Park, Maryland 20742, USA}  

\author{Sarah H. Miller}
\affiliation{Applied Research Laboratory for Intelligence and Security, University of Maryland, College Park, Maryland 20742, USA} 

\author{Alaina M. Green}
\affiliation{Joint Quantum Institute and Department of Physics, University of Maryland, College Park, Maryland 20742, USA} 
\affiliation{National Quantum Laboratory (QLab), University of Maryland, College Park, MD 20742 USA}

\author{Norbert M. Linke}
\affiliation{Joint Quantum Institute and Department of Physics, University of Maryland, College Park, Maryland 20742, USA}
\affiliation{National Quantum Laboratory (QLab), University of Maryland, College Park, MD 20742 USA}
\affiliation{Duke Quantum Center and Department of Physics, Duke University, Durham, North Carolina 27701, USA}

\author{Victor Galitski}
\affiliation{Joint Quantum Institute and Department of Physics, University of Maryland, College Park, Maryland 20742, USA}  
\begin{abstract}

We implement a quantum generalization of a neural network on trapped-ion and IBM superconducting quantum computers to classify MNIST images, a common benchmark in computer vision. The network feedforward involves qubit rotations whose angles depend on the results of measurements in the previous layer. The network is trained via simulation, but inference is performed experimentally on quantum hardware. The classical-to-quantum correspondence is controlled by an interpolation parameter, which is zero in the classical limit. Increasing it introduces quantum uncertainty into the measurements, which is shown to improve network performance at moderate values of the interpolation parameter. We then focus on particular images that fail to be classified by a classical neural network but are detected correctly in the quantum network. For such borderline cases, we observe strong deviations from the simulated behavior. We attribute this to physical noise, which causes the output to fluctuate between nearby minima of the classification energy landscape. Such strong sensitivity to physical noise is absent for clear images. We further benchmark physical noise by inserting additional single-qubit and two-qubit gate pairs into the neural network circuits. Our work provides a springboard toward more complex quantum neural networks on current devices: while the approach is rooted in standard classical machine learning, scaling up such networks may prove classically non-simulable and could offer a route to near-term quantum advantage.
\end{abstract}

\maketitle



There exists a deep analogy between neural networks and spin models in statistical mechanics ~\cite{hopfield1982,amit1985,dotsenko1994,GLASS}. In this correspondence, neurons and weights are analogous to classical spins and couplings, respectively. It is natural to quantize these by promoting them to qubits and gates, respectively~\cite{NAT}. This gives rise to a broad class of quantum neural networks. Much recent work in the field of quantum machine learning has focused on searching for ``quantum advantage''~\cite{REVQML1,REVQML2,Li2015Experimental,REVQML3,REVQML4,REVQML5,REVQML6,EMBED,QNISQ}. A variety of network architectures have been discussed in this context, and while some have been simulated under a variety of assumptions, few have been experimentally run on genuine quantum hardware \cite{QNNCOMPARE,Huang2021,Kong2025}, let alone in the context of benchmarking on, \emph{e.g.}, standard image classification tasks \cite{Cherrat2024}. No matter the implementation, such architectures typically do not allow for smooth interpolation between classical and quantum modes, making comparisons between classical neural networks (NN) and quantum neural networks (QNN) difficult.

In the following, we present an architecture for a quantum neural network, which we dub the Benchmark Quantum Neural Network (BQNN), that has several attractive features: (1) It can be adiabatically tuned between the classical and quantum regime.
(2) The circuit is deployable on any functional quantum computing platform with minimal demands on qubit count, gate fidelity, and connectivity.
(3) Because the classical limit of BQNN reproduces a binary neural network, it can be benchmarked on any reference data set, enabling direct classical-quantum comparisons. 
(4) BQNN is generalizable to more complex hardware set-ups, including those not classically simulable \cite{Bermejo2024,gil-fuster2025}, by introducing partial mid-circuit measurements and feedback, bridging the field of measurement-induced phase transitions~\cite{MIPT} with quantum neural networks.  

After reviewing the theory of this BQNN architecture, we present image classification results on both superconducting and trapped-ion hardware. The latter involves two distinct implementations involving gates driven by microwave and laser interactions as discussed below. The experimental results are largely consistent with the expected behavior informed by classical simulations. In particular, in an intermediate quantum regime where quantum uncertainty in measurements is present but does not fully randomize measurements, the BQNN implemented on real devices outperform the classical counterpart. We also examine how device noise affects performance and find that moderate levels of physical noise do not hinder inference. We even find instances where intentional introduction of noise improves performance, up to a limit. Strong gate noise, modeled here via insertion of pairs of single-qubit and two-qubit gates and their inverse, eventually degrades performance to the level of a random guess. The number of noise gates it takes to see a decrease in performance provides another neural-network-specific metric to characterize gate noise. To our knowledge, this study is the first to compare different hardware implementations of QNNs for benchmarking image classification. 

\begin{figure*}[t]
    \centering
    \subfloat[Mapping from classical NN to BQNN.]{
    \label{fig:nn_diag_updated}
    \includegraphics[width=0.68\textwidth]{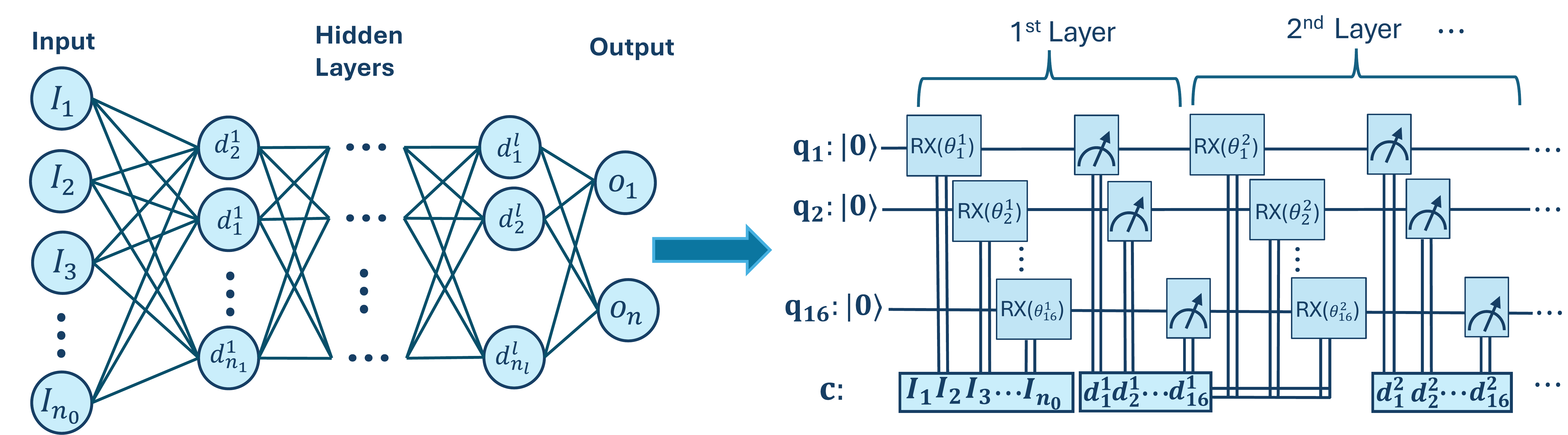}
    }
    \hspace{-0.6em}
    \subfloat[BQNN circuit injecting $U$ and $U^\dagger$ pairs with single- (left) and two-qubit (right) gates.]{
    \label{fig:three_sin_x}
    \includegraphics[width=0.3\textwidth]{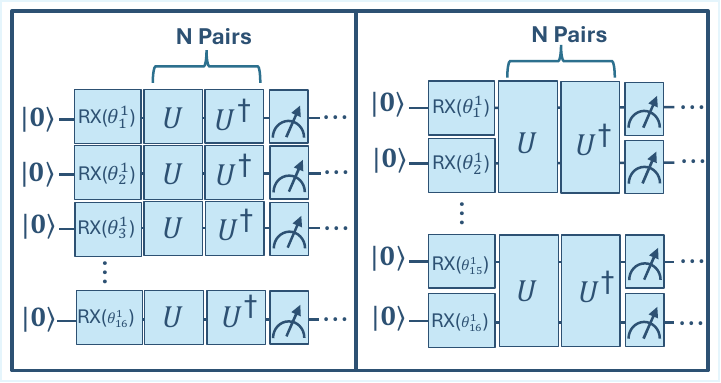}
    }
    \hfill
    \subfloat[Trapped-ion quantum computer with microwave driven gates.]{
    \label{fig:microwave}
    \includegraphics[width=0.3\textwidth]{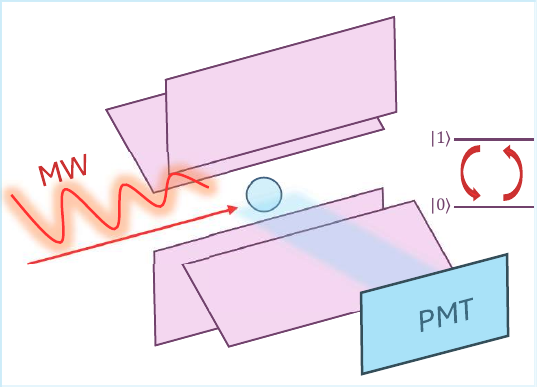}
    }
    \hfill
    \subfloat[Trapped-ion quantum computer with laser driven gates.]{
    \label{fig:laser}
    \includegraphics[width=0.3\textwidth]{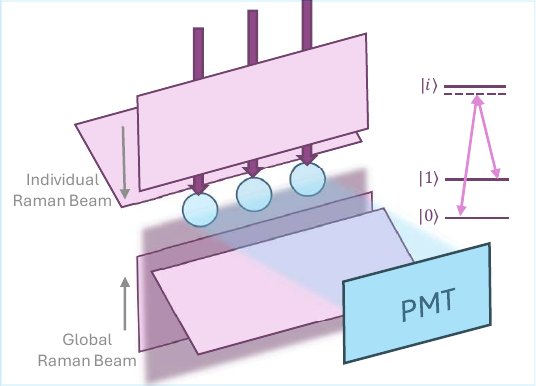}
    }
    \hfill
    \subfloat[Quantum computer based on superconducting qubits.]{
    \label{fig:superconducting}
    \includegraphics[width=0.3\textwidth]{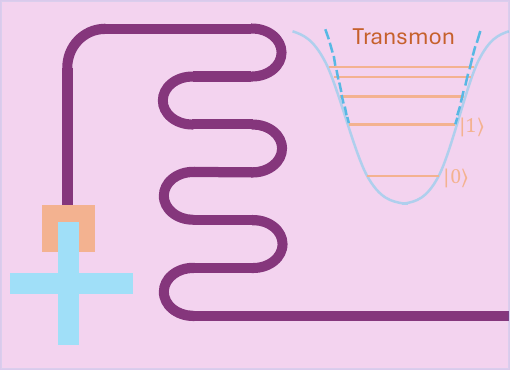}
    }
    \caption{(a) The mapping from the classical binarized multi-layer perceptron network to the quantum circuit implementing the hidden layers of the BQNN. Each step of the forward pass consists of rotating each qubit by an angle controlled by the classical channels, then measuring each qubit. These measurements are the activations which are then passed to the next layer of the network. The networks consist of an input layer, $l$ hidden layers, and an output layer. (b) The BQNN circuit with noise injection by means of inserted $U$ and $U^\dagger$ pairs. (c)-(e) The varieties of quantum hardware on which we run the BQNN: trapped ions with microwave driven gates, trapped ions with laser driven gates, and superconducting qubits.}
    \label{fig:two_graphs}
\end{figure*}

Ref.~\cite{NAT} presents theory and simulation of a quantum partially binarized multilayer perceptron i.e. the BQNN. See Fig.~\ref{fig:nn_diag_updated}. This network may be smoothly tuned between classical and quantum modes of operation. This network is partially binarized in the sense that, while each training weight can take on any real value, the activations of the hidden layers are constrained to be $\pm 1$. For a classical network this is done by using the activation function
\begin{equation}\label{eq:sgn}
    \phi(x)=\text{sgn}(x).
\end{equation}
The activation of the $i^\text{th}$ neuron in the $k^\text{th}$  layer is 
\begin{equation}
    d_i^k=\phi\left((W^k\mathbf d^{k-1})_i\right),
\end{equation}
where $\mathbf d^0=\mathbf I$ is the input vector and each $W^k$ is a matrix of trainable parameters. In this way, information travels through the network layer by layer, in a process called feedforward. The output of the network with $\ell-1$ hidden layers is
\begin{equation}
    \mathbf o=W^\ell\mathbf d^{\ell-1},
\end{equation}
which is then the argument of the loss function during training.

This classical network is then quantized in a straightforward way: binary neurons are promoted to qubits and activation functions are implemented by single-qubit rotations followed by projective measurements. By defining the activations as the results of projective measurements, the activations remain binarized. The circuit which implements the hidden layers is shown in Fig.~\ref{fig:nn_diag_updated}.

To tune  between quantum and classical regimes, we define the rotation angle of  qubit $i$ in  layer $k$ as
\begin{gather}\label{eq:quantum_activation}
    \theta_i^k=\frac\pi 2\begin{cases}
        1-\phi_a\left((W^1\mathbf I)_i\right), & k=1\\
        d_i^{k-1}-\phi_a\left((W^k\mathbf d^{(k-1)})_i\right), & k>1
    \end{cases},\\
    \phi_a(x)=\text{htanh}(x/a)=\begin{cases}
        \text{sgn}(x), & |x|\geq a\\
        x/a, & |x|<a
    \end{cases},
\end{gather}
where $\phi_a$ may be considered a quantum activation function. In the limit $a\rightarrow 0$, $\phi_a$ reduces to the classical activation function Eq.~\ref{eq:sgn}. In this case, each rotation angle is either 0 or $\pm\pi$. This means that, under ideal conditions, the rotation gates will only move the qubits between computational basis states, the behavior of the $a=0$ network is deterministic, and is equivalent to the standard classical neural net. Non-zero values of $a$ enable arbitrary angels of rotation and hence  non-deterministic measurements. We identify $a$ as a parameter describing the ``quantumness" of the network.

 BQNN training was done using stochastic gradient descent with momentum ~\cite{GoodfellowBengioCourville2016GradientBased}. A difficulty in gradient descent arises for both classical and quantized  networks due to the activation function Eq.~(\ref{eq:sgn}) having an undefined first derivative and probabilistic activations correspondingly. To overcome this, a clipped straight-through gradient estimator~\cite{bengio2013estimatingpropagatinggradientsstochastic} was used. I.e., when backpropagating the gradients,  the activations are treated as:
\begin{gather}
    d_i^k=\phi_\text{back}\left((W^k\mathbf d^{k-1})_i\right),\\
    \phi_\text{back}(x)=\text{htanh}(x)=\begin{cases}
        \text{sgn}(x), & |x|\geq 1\\
        x, & |x|<1
    \end{cases}.
\end{gather}
This difference between the forward and backward passes  introduces error into the gradient estimation, but  in practice  yields an effective training procedure~\cite{NIPS2015_3e15cc11}. 

An additional difference between classical and quantum networks arises in inference. While it is only necessary to run inference once for each test datum in the deterministic classical case, in the quantum case, classifications fluctuate from run to run. We take a sample of classifications for a single input and select the outcome by majority vote. Ref.~\cite{NAT} explored both training and inference with a classically-simulated BQNN on the MNIST handwritten digit recognition dataset \cite{NIST1990SD19}, finding that the classification error rate improved over classical-mode results in an intermediate quantum regime where \(0 < a < \mathcal{O}(1)\). This advantage only required on the order of 10 classification samples per image. 

In the present work we still train the network using classical simulation, but perform \textit{classification} on two different sets of hardware i.e. IBM circuit-QED processors based on transmon qubits \cite{QEDcircuit, Cooperpair}, the trapped-ion quantum computer at the Joint Quantum Institute with microwave gates and Raman gates. The network consists of three layers with 16 qubits per layer and is classically trained using a straight-through estimator (STE) on the MNIST dataset.While the circuit in Fig.~\ref{fig:three_sin_x} is constructed on 16 qubits according to the network's architecture, the circuit is separable and hence can be executed on as few as one qubit. The training and classification code can be found in the repositories~\cite{repo2024, repo2025}. Below is a brief description of the hardware.

The ion trap system is described in detail in Ref.~\cite{debnath2016demonstration}. It consists of $^{171}\mathrm{Yb}^+$ ions confined in a linear Paul trap. Each ion possesses two distinct types of quantum states: internal states and motional states. The internal states, used to encode qubits, correspond to hyperfine levels within the electronic ground-state manifold, $^2S_{1/2}$. Specifically, the qubit basis states $|0\rangle$ and $|1\rangle$ are defined as $|F, m_F\rangle=|0, 0\rangle$ and $|1,0\rangle$ respectively. Here, $F$ indicates the total angular momentum including both the nuclear spin and total electronic angular momentum in these hyperfine-split states while $m_F$ is its projection onto the quantization axis. The motional states arise from the quantized collective oscillations of the ion chain within the harmonic potential formed by the radio-frequency and static electric fields of the Paul trap. These quantized vibrational modes serve as a shared quantum bus and play a central role in mediating interactions between qubits in trapped-ion quantum systems.

In this experiment, we employ two trapped-ion gate architectures: microwave-driven single-qubit gates and laser-driven Raman transitions which implement both single-qubit and two-qubit entangling gates. For Raman-based operations, three ions are confined simultaneously, though only the left and right ions were used for quantum gates. Single-qubit gates are implemented using stimulated Raman transitions, driven by two beams derived from a 355 nm mode-locked pulsed laser: one beam globally illuminates the ion chain, while the other is split into tightly focused beams for individual ion addressing. The frequency difference matches the qubit splitting and drives coherent transitions between the qubit states. Entangling gates are implemented by simultaneously applying two bichromatic laser beams—one to each ion— tuned near the red and blue sidebands of a shared motional mode. These beams generate spin-dependent forces that couple the ions’ internal (spin) states through their collective motion, realizing a high-fidelity Molmer–Sorensen interaction.
For microwave gates, a single ion is trapped, and transitions between the states are driven using resonant microwave radiation. This approach enables higher fidelity qubit control independent of the qubit motional state, the latter being a source of error in laser-driven Raman gates.
In both modes, the qubit states are read out by collecting state-dependent fluorescence on a photomultiplier tube  array. 

All superconducting‑qubit experiments are executed on IBM’s commercial Eagle and Heron processors, which host capacitively coupled transmon qubits arranged in a heavy‑hex lattice and controlled with microwave single‑qubit rotations, cross‑resonance two‑qubit gates, and dispersive resonator readout. We access these processors through the IBM Quantum cloud API, submitting circuits  and retrieving results remotely. 

\begin{figure*}[!t]
  \centering
  \begin{minipage}[t]{0.3\textwidth}
    \subfloat[]{\includegraphics[width=\linewidth]{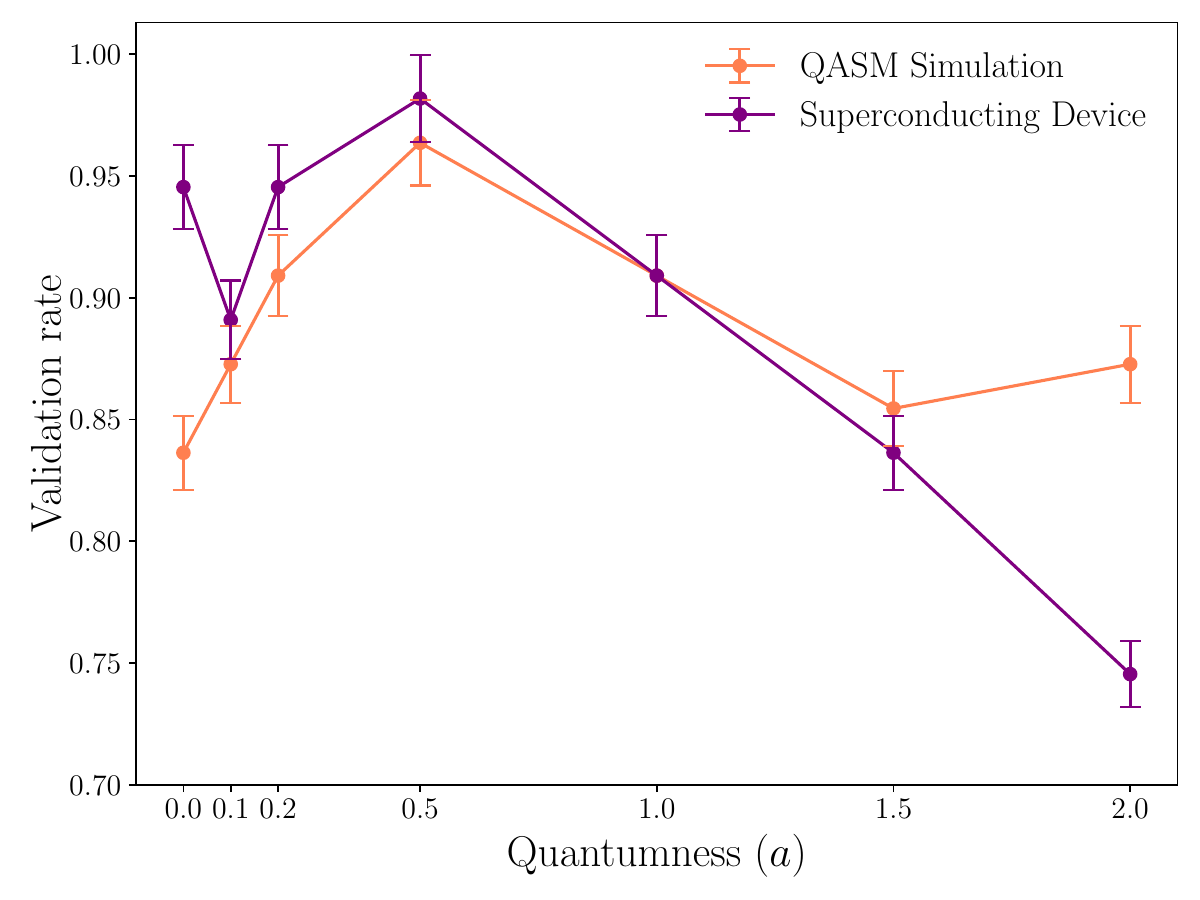}\label{fig:statistics}}\\[1ex]
    \subfloat[]{\includegraphics[width=\linewidth]{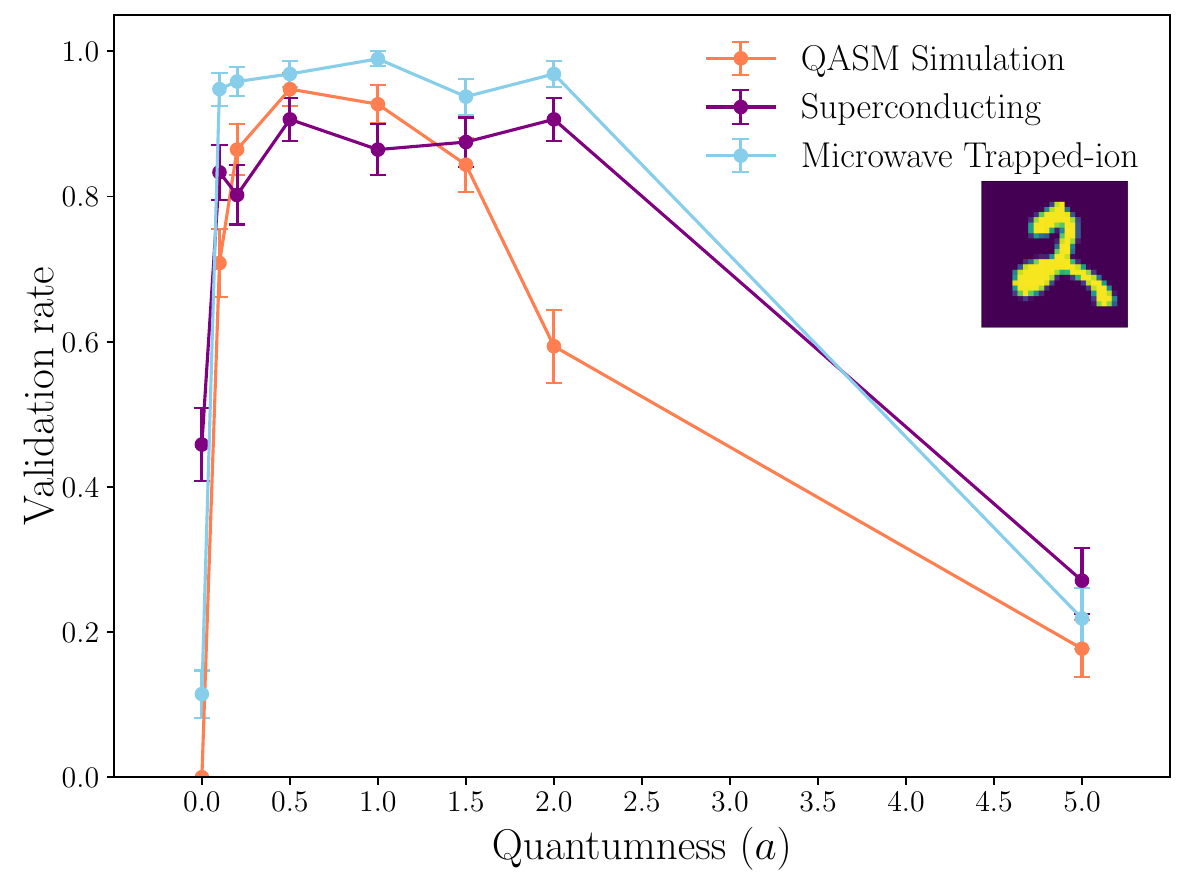}\label{fig:cherry_pick}}
  \end{minipage}
  \begin{minipage}[t]{0.62\textwidth}
    \subfloat[]{\includegraphics[width=\linewidth]{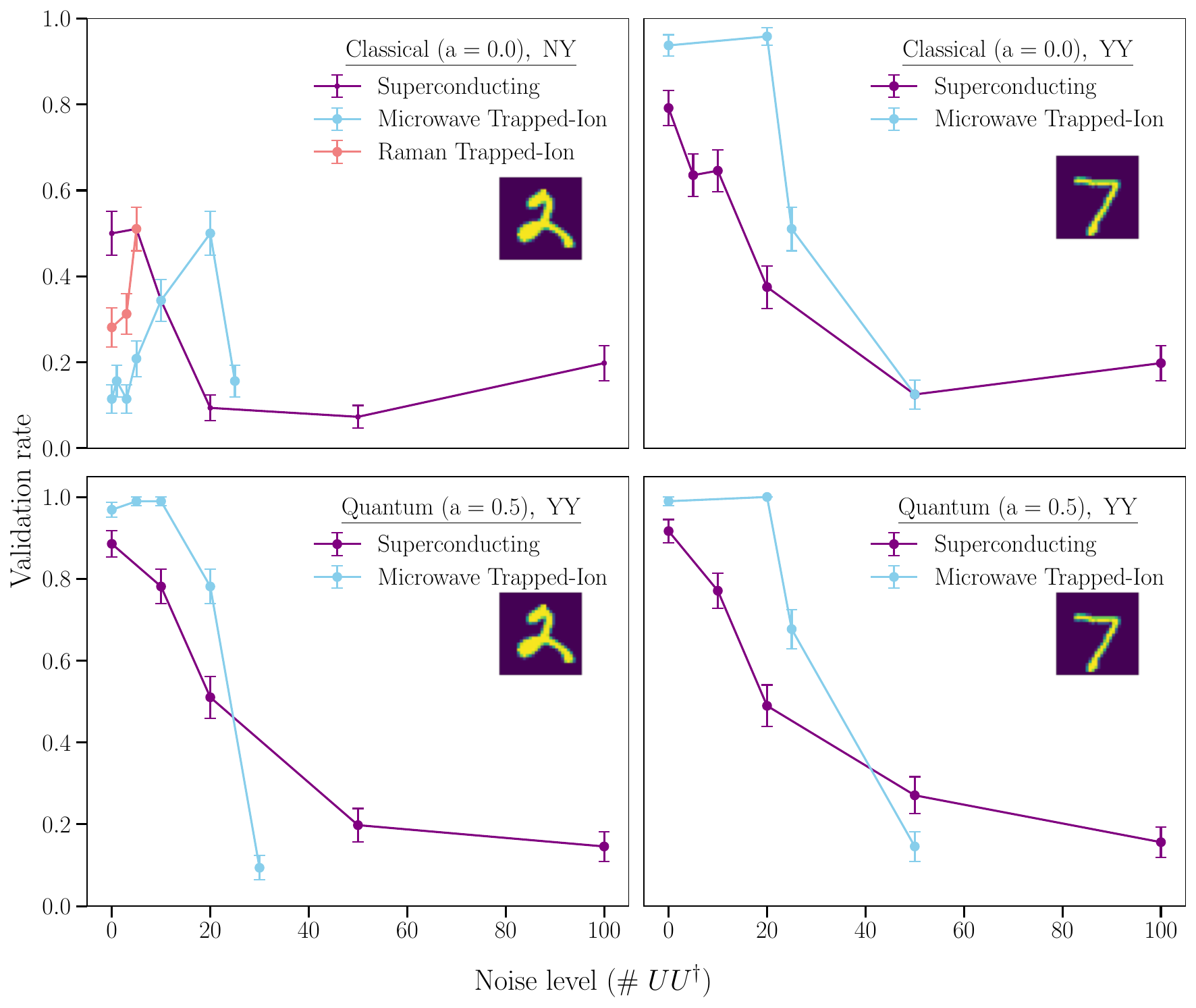}\label{fig:qubit_noise}}
  \end{minipage}
  \caption{(a) Validation rates on 55 randomly selected test images obtained from superconducting hardware and simulation for different values of $a$. For each image, ten shots were taken and the most common outcome was chosen as the predicted class. (b) An NY image that was misclassified at $a=0.0$ shows improved classification accuracy when the quantumness $a$ is increased to some optimal level on the three different platforms, i.e. superconducting, trapped-ion, and classical simulation. (c) For an NY and YY image, taken at either $a=0.0$ (classical) and $a=0.5$ (quantum), noise is controlled by the insertion of a number of single-qubit $UU^\dagger=I$ pairs.}
  \label{fig:3panels}
\end{figure*}

As an initial test of the effectiveness of BQNN, we randomly select and classify 55 images from the MNIST test set, taking 10 samples per image. Due to resource constraints, this initial test is performed only on quantum assembly simulation (QASM) and IBM hardware. The results are depicted in Fig.~\ref{fig:statistics}. Our results broadly agree with the prediction of Ref.~\cite{NAT}. In particular, in an intermediate regime ($a\approx 0.5$) we observe the validation rate of the BQNN greater than that of the analogous classical network.  Surprisingly, the validation rates observed in the experiment slightly exceed those obtained through classical simulation for $a\leq 0.5$. Although the difference is within error bars, its presence can only be attributed to physical noise and demonstrates that noise is not always hindrance to performance. This initial experiment also explicitly verifies that quantum uncertainty in quantum neural nets can play a role similar to  injecting stochastic noise during classical neural‑network training, which is known to act as an effective regularizer and boosts performance~\cite{bishop1995noise}. We explicitly verify a similar behavior here but induced by intrinsically quantum noise.

To further explore the effect of both quantum randomness and physical noise on image classification, we select a representative image (index 6929) that failed to be correctly classified by the classical model ($a=0$) but succeeded in simulations with the introduction of quantum noise ($a>0$). We call such images No-Yes (NY) images, in contrast with Yes-Yes (YY) images, which are correctly classified by both the quantum and classical networks.  Fig.~\ref{fig:cherry_pick} compares results on the NY image between simulation, IBM hardware, and microwave-gate trapped-ions. 

Perhaps the most interesting observation is the difference between the simulated behavior and  experimental results at zero and small values of the quantumness parameter, $a$. At $a=0$, the classical network deterministically fails and so does the idealized quantum simulation of its quantum realization. However, the experimental validation rate at $a=0$ for this image is 10\% for the trapped ion quantum device and as high as 50\% for the IBM device. This behavior too can only be attributed to physical noise. A plausible explanation is that there are two nearby minima in the energy landscape of the trained neural net~\cite{GLASS} - one representing the erroneous reading, which gets selected classically 100\% of the time, and the second being the correct reading.  Measurement noise introduces classical stochasticity and errors, which in this case switch to the correct classification in a subset of runs. We note that we have not observed such sensitivity to physical noise at $a=0$ for clear YY images (where presumably there is a single deep minimum insensitive to small perturbations). This simple observation suggests an interesting use case for noisy quantum neural networks as a way to flag ambiguous data entries. For the MNIST dataset, the data quality is relatively obvious from visual inspection, but this is not necessarily the case for more complex datasets. 

The NY images are the source of the boost of performance upon quantization exemplified in Fig.~\ref{fig:statistics}. We explore the validation rate as a function of quantumness, $a$, for the particular chosen image in Fig.~\ref{fig:cherry_pick}. We observe a steep rise in the validation rate for this image to nearly 100\% on the trapped ion device and about 90\% on the IBM quantum computer. When $a$ becomes sufficiently large, it introduces nearly random rotations by producing qubit cat-states with almost equal probabilities for the two computational basis states. This predictably hinders classification performance and we observe validation rates dropping with with $a$. In the  $a \to \infty$ limit classifications becomes uniformly random, and the validation rate converges to 0.1 in accordance with  Eq.~\eqref{eq:quantum_activation}. 

The validation rate inferred from both quantum platforms shows a subtle maximum at $a \approx 0.5$, but otherwise exhibits a plateau in an intermediate quantum regime before eventual performance decay. Owing to its sensitivity to physical noise, the $a=0$ and optimal $a$ regimes of the BQNN allow for a loose comparison of the physical noise between platforms. In particular, it is plausible that for these low-depth circuits, physical errors are the largest portion of the error. Hence, devices with relatively high measurement errors, such as superconducting devices, will see a greater noise-induced rise in validation rate~\cite{benchmark}. In order to serve as a more stringent benchmark of quantum devices, entanglement and mid-circuit measurement will need to be incorporated into the BQNN.

We further modify the BQNN to benchmark the effects of gate noise on feedforward dynamics. Specifically, $n$ pairs of {native} unitary gates $UU^\dagger = I$ are injected into the circuit after each layer. While in an ideal system, they have no effect, their application  in a real device  enhances bit-flip, phase-flip and other types of noise in the case of single-qubit pairs  and introduce ``parasitic entanglement'' and  crosstalk for pairs of time-reversed entangling gates.



\begin{figure}[!t]
    \includegraphics[width=1.0\linewidth]{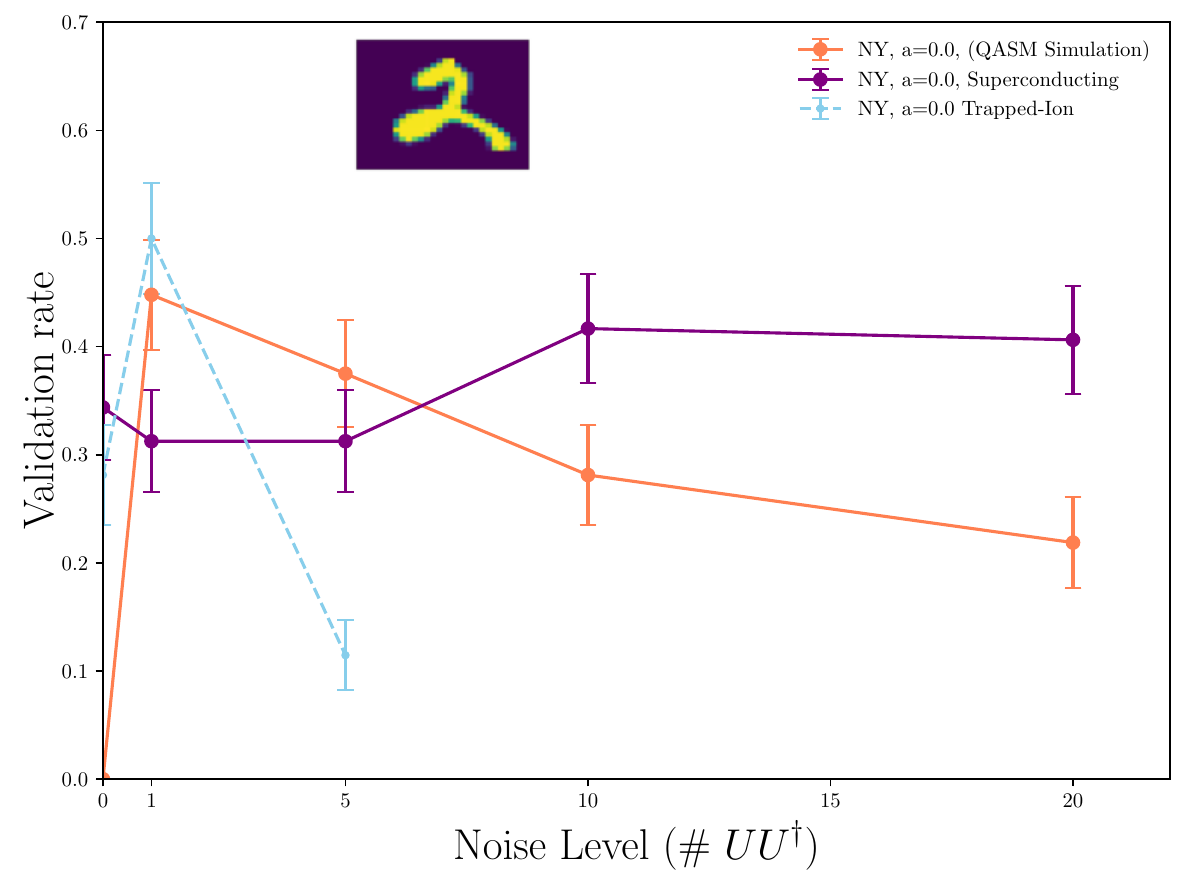}
    \caption{For an NY image at $a=0.0$, noise is controlled by the insertion of a number of two-qubit $UU^\dagger=I$ pairs.}
    \label{fig:two_qubit_noise}
\end{figure}

Fig.~\ref{fig:qubit_noise} shows results of our single-qubit noise experiments. In the classical regime, we observe for the trapped ion device that even modest amounts of injected noise lead to a clear jump in NY-image accuracy at around 20 noise-gate pairs. We do not see any notable improvement of performance on the YY image with injected noise, which further validates the heuristic picture of two close by minima for the borderline (ambiguous) NY data entries versus a single deep minimum for the clear YY images. This behavior carries over into the moderately quantum regime, where the combination of quantum randomness and added single-qubit noise has only a minor impact on validation rates. By contrast, the superconducting system does not exhibit any enhancement in performance on the NY images with injected noise suggesting that the native physical measurement noise is already saturated or above the threshold to be beneficial for the BQNN performance. We see the  per-image accuracy decline almost monotonically as noise increases.

The final test uses native two-qubit gate pairs and results are depicted in Fig.~\ref{fig:two_qubit_noise}. For the trapped-ion experiment it was found that after the introduction of a \textit{single} pair of two-qubit gates, the per-image classification-rate shot up over $40\%$ before deteriorating to random 10\% for five gate pairs. The superconducting platform was more resilient to this entanglement-noise and even showed a small improvement in performance with the addition of further pairs, a behavior not consistent with our intuition or simulations.

All in all, our experimental results suggest the possibility that certain types of physical noise, usually thought detrimental to performance of quantum computers, can be beneficial for certain quantum machine learning tasks. While such instances at the inference stage may be accidental or task-specific, we expect a more systematic advantage when training quantum networks with small amounts of background physical noise.

The main obstacle to generalizing this architecture to quantum networks that cannot be efficiently simulated classically, and thus with potential for genuine quantum advantage, is the challenge of implementing mid-circuit measurements and feedback. Without measurements, any circuit is a subset of a unitary matrix which lacks non-linear input. On the other hand, with complete measurement in every layer the quantum state collapses and entanglement is disrupted, hence leading to circuits which can always be simulated classically. The only alternative is entangled (e.g., brick-wall) circuits with partial mid-circuit measurements. This corresponds to circuits designed for measurement-induced phase transitions with the difference being that in neural nets, gates applied in every layer are a function of the partial measurement in the previous layer. Such measurement-induced quantum neural nets can be trained using reinforcement learning to aid in architecture search, which will be reported elsewhere.

\begin{acknowledgments}

This work is supported by a collaboration between the US DOE and other Agencies. This material is based upon work supported by the U.S. Department of Energy, Office of Science, National Quantum Information Science Research Centers, Quantum Systems Accelerator under Award No. DE-FOA-0002253 (XL, AMG, NML) and Basic Energy Sciences under Award No. DE-SC0001911 (VG, RB).
Additional support is acknowledged from the Army Research Office under Grant Number W911NF-23-1-0241 (DJH) and from the National Science Foundation STAQ project No. PHY-2325080 (NML).
We acknowledge the use of IBM Quantum Credits for this work. The views expressed are those of the authors, and do not reflect the official policy or position of IBM or the IBM Quantum team. 

\end{acknowledgments}

\bibliographystyle{apsrev4-1}
\bibliography{refs}  

\begin{thebibliography}{32}%
\makeatletter
\providecommand \@ifxundefined [1]{%
 \@ifx{#1\undefined}
}%
\providecommand \@ifnum [1]{%
 \ifnum #1\expandafter \@firstoftwo
 \else \expandafter \@secondoftwo
 \fi
}%
\providecommand \@ifx [1]{%
 \ifx #1\expandafter \@firstoftwo
 \else \expandafter \@secondoftwo
 \fi
}%
\providecommand \natexlab [1]{#1}%
\providecommand \enquote  [1]{``#1''}%
\providecommand \bibnamefont  [1]{#1}%
\providecommand \bibfnamefont [1]{#1}%
\providecommand \citenamefont [1]{#1}%
\providecommand \href@noop [0]{\@secondoftwo}%
\providecommand \href [0]{\begingroup \@sanitize@url \@href}%
\providecommand \@href[1]{\@@startlink{#1}\@@href}%
\providecommand \@@href[1]{\endgroup#1\@@endlink}%
\providecommand \@sanitize@url [0]{\catcode `\\12\catcode `\$12\catcode `\&12\catcode `\#12\catcode `\^12\catcode `\_12\catcode `\%12\relax}%
\providecommand \@@startlink[1]{}%
\providecommand \@@endlink[0]{}%
\providecommand \url  [0]{\begingroup\@sanitize@url \@url }%
\providecommand \@url [1]{\endgroup\@href {#1}{\urlprefix }}%
\providecommand \urlprefix  [0]{URL }%
\providecommand \Eprint [0]{\href }%
\providecommand \doibase [0]{http://dx.doi.org/}%
\providecommand \selectlanguage [0]{\@gobble}%
\providecommand \bibinfo  [0]{\@secondoftwo}%
\providecommand \bibfield  [0]{\@secondoftwo}%
\providecommand \translation [1]{[#1]}%
\providecommand \BibitemOpen [0]{}%
\providecommand \bibitemStop [0]{}%
\providecommand \bibitemNoStop [0]{.\EOS\space}%
\providecommand \EOS [0]{\spacefactor3000\relax}%
\providecommand \BibitemShut  [1]{\csname bibitem#1\endcsname}%
\let\auto@bib@innerbib\@empty
\bibitem [{\citenamefont {Hopfield}(1982)}]{hopfield1982}%
  \BibitemOpen
  \bibfield  {author} {\bibinfo {author} {\bibfnamefont {J.~J.}\ \bibnamefont {Hopfield}},\ }\href {\doibase 10.1073/pnas.79.8.2554} {\bibfield  {journal} {\bibinfo  {journal} {Proceedings of the National Academy of Sciences}\ }\textbf {\bibinfo {volume} {79}},\ \bibinfo {pages} {2554} (\bibinfo {year} {1982})}\BibitemShut {NoStop}%
\bibitem [{\citenamefont {Amit}\ \emph {et~al.}(1985)\citenamefont {Amit}, \citenamefont {Gutfreund},\ and\ \citenamefont {Sompolinsky}}]{amit1985}%
  \BibitemOpen
  \bibfield  {author} {\bibinfo {author} {\bibfnamefont {D.~J.}\ \bibnamefont {Amit}}, \bibinfo {author} {\bibfnamefont {H.}~\bibnamefont {Gutfreund}}, \ and\ \bibinfo {author} {\bibfnamefont {H.}~\bibnamefont {Sompolinsky}},\ }\href {\doibase 10.1103/PhysRevA.32.1007} {\bibfield  {journal} {\bibinfo  {journal} {Phys. Rev. A}\ }\textbf {\bibinfo {volume} {32}},\ \bibinfo {pages} {1007} (\bibinfo {year} {1985})}\BibitemShut {NoStop}%
\bibitem [{\citenamefont {Dotsenko}(1994)}]{dotsenko1994}%
  \BibitemOpen
  \bibfield  {author} {\bibinfo {author} {\bibfnamefont {V.}~\bibnamefont {Dotsenko}},\ }\href@noop {} {\emph {\bibinfo {title} {An introduction to the theory of spin glasses and neural networks}}},\ \bibinfo {series} {World Scientific lecture notes in physics}, Vol.~\bibinfo {volume} {55}\ (\bibinfo  {publisher} {World Scientific},\ \bibinfo {address} {Singapore},\ \bibinfo {year} {1994})\BibitemShut {NoStop}%
\bibitem [{\citenamefont {Barney}\ \emph {et~al.}(2024)\citenamefont {Barney}, \citenamefont {Winer},\ and\ \citenamefont {Galitski}}]{GLASS}%
  \BibitemOpen
  \bibfield  {author} {\bibinfo {author} {\bibfnamefont {R.}~\bibnamefont {Barney}}, \bibinfo {author} {\bibfnamefont {M.}~\bibnamefont {Winer}}, \ and\ \bibinfo {author} {\bibfnamefont {V.}~\bibnamefont {Galitski}},\ }\href {https://arxiv.org/abs/2408.06421} {\enquote {\bibinfo {title} {Neural networks as spin models: From glass to hidden order through training},}\ } (\bibinfo {year} {2024}),\ \Eprint {http://arxiv.org/abs/2408.06421} {arXiv:2408.06421 [cond-mat.dis-nn]} \BibitemShut {NoStop}%
\bibitem [{\citenamefont {Barney}\ \emph {et~al.}(2025)\citenamefont {Barney}, \citenamefont {Lakhdar-Hamina},\ and\ \citenamefont {Galitski}}]{NAT}%
  \BibitemOpen
  \bibfield  {author} {\bibinfo {author} {\bibfnamefont {R.}~\bibnamefont {Barney}}, \bibinfo {author} {\bibfnamefont {D.}~\bibnamefont {Lakhdar-Hamina}}, \ and\ \bibinfo {author} {\bibfnamefont {V.}~\bibnamefont {Galitski}},\ }\href {https://arxiv.org/abs/2503.15482} {\enquote {\bibinfo {title} {Natural quantization of neural networks},}\ } (\bibinfo {year} {2025}),\ \Eprint {http://arxiv.org/abs/2503.15482} {arXiv:2503.15482 [quant-ph]} \BibitemShut {NoStop}%
\bibitem [{\citenamefont {Pudenz}\ and\ \citenamefont {Lidar}(2013)}]{REVQML1}%
  \BibitemOpen
  \bibfield  {author} {\bibinfo {author} {\bibfnamefont {K.~L.}\ \bibnamefont {Pudenz}}\ and\ \bibinfo {author} {\bibfnamefont {D.~A.}\ \bibnamefont {Lidar}},\ }\href {\doibase 10.1007/s11128-012-0506-4} {\bibfield  {journal} {\bibinfo  {journal} {Quantum Information Processing}\ }\textbf {\bibinfo {volume} {12}},\ \bibinfo {pages} {2027} (\bibinfo {year} {2013})}\BibitemShut {NoStop}%
\bibitem [{\citenamefont {Lloyd}\ \emph {et~al.}(2013)\citenamefont {Lloyd}, \citenamefont {Mohseni},\ and\ \citenamefont {Rebentrost}}]{REVQML2}%
  \BibitemOpen
  \bibfield  {author} {\bibinfo {author} {\bibfnamefont {S.}~\bibnamefont {Lloyd}}, \bibinfo {author} {\bibfnamefont {M.}~\bibnamefont {Mohseni}}, \ and\ \bibinfo {author} {\bibfnamefont {P.}~\bibnamefont {Rebentrost}},\ }\href {https://arxiv.org/abs/1307.0411} {\enquote {\bibinfo {title} {Quantum algorithms for supervised and unsupervised machine learning},}\ } (\bibinfo {year} {2013}),\ \Eprint {http://arxiv.org/abs/1307.0411} {arXiv:1307.0411 [quant-ph]} \BibitemShut {NoStop}%
\bibitem [{\citenamefont {Li}\ \emph {et~al.}(2015)\citenamefont {Li}, \citenamefont {Liu}, \citenamefont {Xu},\ and\ \citenamefont {Du}}]{Li2015Experimental}%
  \BibitemOpen
  \bibfield  {author} {\bibinfo {author} {\bibfnamefont {Z.}~\bibnamefont {Li}}, \bibinfo {author} {\bibfnamefont {X.}~\bibnamefont {Liu}}, \bibinfo {author} {\bibfnamefont {N.}~\bibnamefont {Xu}}, \ and\ \bibinfo {author} {\bibfnamefont {J.}~\bibnamefont {Du}},\ }\href {\doibase 10.1103/PhysRevLett.114.140504} {\bibfield  {journal} {\bibinfo  {journal} {Phys. Rev. Lett.}\ }\textbf {\bibinfo {volume} {114}},\ \bibinfo {pages} {140504} (\bibinfo {year} {2015})}\BibitemShut {NoStop}%
\bibitem [{\citenamefont {Ciliberto}\ \emph {et~al.}(2018)\citenamefont {Ciliberto}, \citenamefont {Herbster}, \citenamefont {Ialongo}, \citenamefont {Pontil}, \citenamefont {Rocchetto}, \citenamefont {Severini},\ and\ \citenamefont {Wossnig}}]{REVQML3}%
  \BibitemOpen
  \bibfield  {author} {\bibinfo {author} {\bibfnamefont {C.}~\bibnamefont {Ciliberto}}, \bibinfo {author} {\bibfnamefont {M.}~\bibnamefont {Herbster}}, \bibinfo {author} {\bibfnamefont {A.~D.}\ \bibnamefont {Ialongo}}, \bibinfo {author} {\bibfnamefont {M.}~\bibnamefont {Pontil}}, \bibinfo {author} {\bibfnamefont {A.}~\bibnamefont {Rocchetto}}, \bibinfo {author} {\bibfnamefont {S.}~\bibnamefont {Severini}}, \ and\ \bibinfo {author} {\bibfnamefont {L.}~\bibnamefont {Wossnig}},\ }\href {\doibase https://doi.org/10.1098/rspa.2017.0551} {\bibfield  {journal} {\bibinfo  {journal} {Proc. Math. Phys. Eng. Sci.}\ }\textbf {\bibinfo {volume} {474}},\ \bibinfo {pages} {20170551} (\bibinfo {year} {2018})}\BibitemShut {NoStop}%
\bibitem [{\citenamefont {Schuld}\ \emph {et~al.}(2015)\citenamefont {Schuld}, \citenamefont {Sinayskiy},\ and\ \citenamefont {Petruccione}}]{REVQML4}%
  \BibitemOpen
  \bibfield  {author} {\bibinfo {author} {\bibfnamefont {M.}~\bibnamefont {Schuld}}, \bibinfo {author} {\bibfnamefont {I.}~\bibnamefont {Sinayskiy}}, \ and\ \bibinfo {author} {\bibfnamefont {F.}~\bibnamefont {Petruccione}},\ }\href {\doibase 10.1080/00107514.2014.964942} {\bibfield  {journal} {\bibinfo  {journal} {Contemporary Physics}\ }\textbf {\bibinfo {volume} {56}},\ \bibinfo {pages} {172} (\bibinfo {year} {2015})}\BibitemShut {NoStop}%
\bibitem [{\citenamefont {Biamonte}\ \emph {et~al.}(2017)\citenamefont {Biamonte}, \citenamefont {Wittek}, \citenamefont {Pancotti}, \citenamefont {Rebentrost}, \citenamefont {Wiebe},\ and\ \citenamefont {Lloyd}}]{REVQML5}%
  \BibitemOpen
  \bibfield  {author} {\bibinfo {author} {\bibfnamefont {J.}~\bibnamefont {Biamonte}}, \bibinfo {author} {\bibfnamefont {P.}~\bibnamefont {Wittek}}, \bibinfo {author} {\bibfnamefont {N.}~\bibnamefont {Pancotti}}, \bibinfo {author} {\bibfnamefont {P.}~\bibnamefont {Rebentrost}}, \bibinfo {author} {\bibfnamefont {N.}~\bibnamefont {Wiebe}}, \ and\ \bibinfo {author} {\bibfnamefont {S.}~\bibnamefont {Lloyd}},\ }\href {\doibase https://doi.org/10.1038/nature23474} {\bibfield  {journal} {\bibinfo  {journal} {Nature}\ }\textbf {\bibinfo {volume} {549}},\ \bibinfo {pages} {195–202} (\bibinfo {year} {2017})}\BibitemShut {NoStop}%
\bibitem [{\citenamefont {Wang}\ and\ \citenamefont {Liu}(2024)}]{REVQML6}%
  \BibitemOpen
  \bibfield  {author} {\bibinfo {author} {\bibfnamefont {Y.}~\bibnamefont {Wang}}\ and\ \bibinfo {author} {\bibfnamefont {J.}~\bibnamefont {Liu}},\ }\href {\doibase 10.1088/1361-6633/ad7f69} {\bibfield  {journal} {\bibinfo  {journal} {Reports on Progress in Physics}\ }\textbf {\bibinfo {volume} {87}},\ \bibinfo {pages} {116402} (\bibinfo {year} {2024})}\BibitemShut {NoStop}%
\bibitem [{\citenamefont {Lloyd}\ \emph {et~al.}(2020)\citenamefont {Lloyd}, \citenamefont {Schuld}, \citenamefont {Ijaz}, \citenamefont {Izaac},\ and\ \citenamefont {Killoran}}]{EMBED}%
  \BibitemOpen
  \bibfield  {author} {\bibinfo {author} {\bibfnamefont {S.}~\bibnamefont {Lloyd}}, \bibinfo {author} {\bibfnamefont {M.}~\bibnamefont {Schuld}}, \bibinfo {author} {\bibfnamefont {A.}~\bibnamefont {Ijaz}}, \bibinfo {author} {\bibfnamefont {J.}~\bibnamefont {Izaac}}, \ and\ \bibinfo {author} {\bibfnamefont {N.}~\bibnamefont {Killoran}},\ }\href {\doibase 10.48550/arXiv.2001.03622} {\enquote {\bibinfo {title} {Quantum embeddings for machine learning},}\ } (\bibinfo {year} {2020})\BibitemShut {NoStop}%
\bibitem [{\citenamefont {Preskill}(2018)}]{QNISQ}%
  \BibitemOpen
  \bibfield  {author} {\bibinfo {author} {\bibfnamefont {J.}~\bibnamefont {Preskill}},\ }\href {\doibase https://doi.org/10.22331/q-2018-08-06-79} {\bibfield  {journal} {\bibinfo  {journal} {Quantum}\ }\textbf {\bibinfo {volume} {2}},\ \bibinfo {pages} {79} (\bibinfo {year} {2018})}\BibitemShut {NoStop}%
\bibitem [{\citenamefont {Abbas}\ \emph {et~al.}(2021)\citenamefont {Abbas}, \citenamefont {Sutter}, \citenamefont {Zoufal}, \citenamefont {Lucchi}, \citenamefont {Figalli},\ and\ \citenamefont {Woerner}}]{QNNCOMPARE}%
  \BibitemOpen
  \bibfield  {author} {\bibinfo {author} {\bibfnamefont {A.}~\bibnamefont {Abbas}}, \bibinfo {author} {\bibfnamefont {D.}~\bibnamefont {Sutter}}, \bibinfo {author} {\bibfnamefont {C.}~\bibnamefont {Zoufal}}, \bibinfo {author} {\bibfnamefont {A.}~\bibnamefont {Lucchi}}, \bibinfo {author} {\bibfnamefont {A.}~\bibnamefont {Figalli}}, \ and\ \bibinfo {author} {\bibfnamefont {S.}~\bibnamefont {Woerner}},\ }\href {\doibase 10.1038/s43588-021-00084-1} {\bibfield  {journal} {\bibinfo  {journal} {Nature Computational Science}\ }\textbf {\bibinfo {volume} {1}},\ \bibinfo {pages} {403} (\bibinfo {year} {2021})}\BibitemShut {NoStop}%
\bibitem [{\citenamefont {Huang}\ \emph {et~al.}(2021)\citenamefont {Huang}, \citenamefont {Du}, \citenamefont {Gong}, \citenamefont {Zhao}, \citenamefont {Wu}, \citenamefont {Wang}, \citenamefont {Li}, \citenamefont {Liang}, \citenamefont {Lin}, \citenamefont {Xu}, \citenamefont {Yang}, \citenamefont {Liu}, \citenamefont {Hsieh}, \citenamefont {Deng}, \citenamefont {Rong}, \citenamefont {Peng}, \citenamefont {Lu}, \citenamefont {Chen}, \citenamefont {Tao}, \citenamefont {Zhu},\ and\ \citenamefont {Pan}}]{Huang2021}%
  \BibitemOpen
  \bibfield  {author} {\bibinfo {author} {\bibfnamefont {H.-L.}\ \bibnamefont {Huang}}, \bibinfo {author} {\bibfnamefont {Y.}~\bibnamefont {Du}}, \bibinfo {author} {\bibfnamefont {M.}~\bibnamefont {Gong}}, \bibinfo {author} {\bibfnamefont {Y.}~\bibnamefont {Zhao}}, \bibinfo {author} {\bibfnamefont {Y.}~\bibnamefont {Wu}}, \bibinfo {author} {\bibfnamefont {C.}~\bibnamefont {Wang}}, \bibinfo {author} {\bibfnamefont {S.}~\bibnamefont {Li}}, \bibinfo {author} {\bibfnamefont {F.}~\bibnamefont {Liang}}, \bibinfo {author} {\bibfnamefont {J.}~\bibnamefont {Lin}}, \bibinfo {author} {\bibfnamefont {Y.}~\bibnamefont {Xu}}, \bibinfo {author} {\bibfnamefont {R.}~\bibnamefont {Yang}}, \bibinfo {author} {\bibfnamefont {T.}~\bibnamefont {Liu}}, \bibinfo {author} {\bibfnamefont {M.-H.}\ \bibnamefont {Hsieh}}, \bibinfo {author} {\bibfnamefont {H.}~\bibnamefont {Deng}}, \bibinfo {author} {\bibfnamefont {H.}~\bibnamefont {Rong}}, \bibinfo {author} {\bibfnamefont {C.-Z.}\ \bibnamefont {Peng}}, \bibinfo {author} {\bibfnamefont
  {C.-Y.}\ \bibnamefont {Lu}}, \bibinfo {author} {\bibfnamefont {Y.-A.}\ \bibnamefont {Chen}}, \bibinfo {author} {\bibfnamefont {D.}~\bibnamefont {Tao}}, \bibinfo {author} {\bibfnamefont {X.}~\bibnamefont {Zhu}}, \ and\ \bibinfo {author} {\bibfnamefont {J.-W.}\ \bibnamefont {Pan}},\ }\href {\doibase 10.1103/PhysRevApplied.16.024051} {\bibfield  {journal} {\bibinfo  {journal} {Phys. Rev. Appl.}\ }\textbf {\bibinfo {volume} {16}},\ \bibinfo {pages} {024051} (\bibinfo {year} {2021})}\BibitemShut {NoStop}%
\bibitem [{\citenamefont {Kong}\ \emph {et~al.}(2025)\citenamefont {Kong}, \citenamefont {Li}, \citenamefont {Chen}, \citenamefont {Xue}, \citenamefont {Xu}, \citenamefont {Liu}, \citenamefont {Wu}, \citenamefont {Fang}, \citenamefont {Fang}, \citenamefont {Chen}, \citenamefont {Yang}, \citenamefont {Dou},\ and\ \citenamefont {Guo}}]{Kong2025}%
  \BibitemOpen
  \bibfield  {author} {\bibinfo {author} {\bibfnamefont {X.}~\bibnamefont {Kong}}, \bibinfo {author} {\bibfnamefont {L.}~\bibnamefont {Li}}, \bibinfo {author} {\bibfnamefont {Z.}~\bibnamefont {Chen}}, \bibinfo {author} {\bibfnamefont {C.}~\bibnamefont {Xue}}, \bibinfo {author} {\bibfnamefont {X.}~\bibnamefont {Xu}}, \bibinfo {author} {\bibfnamefont {H.}~\bibnamefont {Liu}}, \bibinfo {author} {\bibfnamefont {Y.}~\bibnamefont {Wu}}, \bibinfo {author} {\bibfnamefont {Y.}~\bibnamefont {Fang}}, \bibinfo {author} {\bibfnamefont {H.}~\bibnamefont {Fang}}, \bibinfo {author} {\bibfnamefont {K.}~\bibnamefont {Chen}}, \bibinfo {author} {\bibfnamefont {Y.}~\bibnamefont {Yang}}, \bibinfo {author} {\bibfnamefont {M.}~\bibnamefont {Dou}}, \ and\ \bibinfo {author} {\bibfnamefont {G.}~\bibnamefont {Guo}},\ }\href {https://arxiv.org/abs/2503.12790} {\enquote {\bibinfo {title} {Quantum-enhanced llm efficient fine tuning},}\ } (\bibinfo {year} {2025}),\ \Eprint {http://arxiv.org/abs/2503.12790} {arXiv:2503.12790 [quant-ph]}
  \BibitemShut {NoStop}%
\bibitem [{\citenamefont {Cherrat}\ \emph {et~al.}(2024)\citenamefont {Cherrat}, \citenamefont {Kerenidis}, \citenamefont {Mathur}, \citenamefont {Landman}, \citenamefont {Strahm},\ and\ \citenamefont {Li}}]{Cherrat2024}%
  \BibitemOpen
  \bibfield  {author} {\bibinfo {author} {\bibfnamefont {E.~A.}\ \bibnamefont {Cherrat}}, \bibinfo {author} {\bibfnamefont {I.}~\bibnamefont {Kerenidis}}, \bibinfo {author} {\bibfnamefont {N.}~\bibnamefont {Mathur}}, \bibinfo {author} {\bibfnamefont {J.}~\bibnamefont {Landman}}, \bibinfo {author} {\bibfnamefont {M.}~\bibnamefont {Strahm}}, \ and\ \bibinfo {author} {\bibfnamefont {Y.~Y.}\ \bibnamefont {Li}},\ }\href {\doibase 10.22331/q-2024-02-22-1265} {\bibfield  {journal} {\bibinfo  {journal} {{Quantum}}\ }\textbf {\bibinfo {volume} {8}},\ \bibinfo {pages} {1265} (\bibinfo {year} {2024})}\BibitemShut {NoStop}%
\bibitem [{\citenamefont {Bermejo}\ \emph {et~al.}(2024)\citenamefont {Bermejo}, \citenamefont {Braccia}, \citenamefont {Rudolph}, \citenamefont {Holmes}, \citenamefont {Cincio},\ and\ \citenamefont {Cerezo}}]{Bermejo2024}%
  \BibitemOpen
  \bibfield  {author} {\bibinfo {author} {\bibfnamefont {P.}~\bibnamefont {Bermejo}}, \bibinfo {author} {\bibfnamefont {P.}~\bibnamefont {Braccia}}, \bibinfo {author} {\bibfnamefont {M.~S.}\ \bibnamefont {Rudolph}}, \bibinfo {author} {\bibfnamefont {Z.}~\bibnamefont {Holmes}}, \bibinfo {author} {\bibfnamefont {L.}~\bibnamefont {Cincio}}, \ and\ \bibinfo {author} {\bibfnamefont {M.}~\bibnamefont {Cerezo}},\ }\href {https://arxiv.org/abs/2408.12739} {\enquote {\bibinfo {title} {Quantum convolutional neural networks are (effectively) classically simulable},}\ } (\bibinfo {year} {2024}),\ \Eprint {http://arxiv.org/abs/2408.12739} {arXiv:2408.12739 [quant-ph]} \BibitemShut {NoStop}%
\bibitem [{\citenamefont {Gil-Fuster}\ \emph {et~al.}(2025)\citenamefont {Gil-Fuster}, \citenamefont {Gyurik}, \citenamefont {Perez-Salinas},\ and\ \citenamefont {Dunjko}}]{gil-fuster2025}%
  \BibitemOpen
  \bibfield  {author} {\bibinfo {author} {\bibfnamefont {E.}~\bibnamefont {Gil-Fuster}}, \bibinfo {author} {\bibfnamefont {C.}~\bibnamefont {Gyurik}}, \bibinfo {author} {\bibfnamefont {A.}~\bibnamefont {Perez-Salinas}}, \ and\ \bibinfo {author} {\bibfnamefont {V.}~\bibnamefont {Dunjko}},\ }in\ \href {https://openreview.net/forum?id=TdqaZbQvdi} {\emph {\bibinfo {booktitle} {The Thirteenth International Conference on Learning Representations}}}\ (\bibinfo {year} {2025})\BibitemShut {NoStop}%
\bibitem [{\citenamefont {Li}\ \emph {et~al.}(2019)\citenamefont {Li}, \citenamefont {Chen},\ and\ \citenamefont {Fisher}}]{MIPT}%
  \BibitemOpen
  \bibfield  {author} {\bibinfo {author} {\bibfnamefont {Y.}~\bibnamefont {Li}}, \bibinfo {author} {\bibfnamefont {X.}~\bibnamefont {Chen}}, \ and\ \bibinfo {author} {\bibfnamefont {M.~P.~A.}\ \bibnamefont {Fisher}},\ }\href {\doibase 10.1103/PhysRevB.100.134306} {\bibfield  {journal} {\bibinfo  {journal} {Phys. Rev. B}\ }\textbf {\bibinfo {volume} {100}},\ \bibinfo {pages} {134306} (\bibinfo {year} {2019})}\BibitemShut {NoStop}%
\bibitem [{\citenamefont {Goodfellow}\ \emph {et~al.}(2016)\citenamefont {Goodfellow}, \citenamefont {Bengio},\ and\ \citenamefont {Courville}}]{GoodfellowBengioCourville2016GradientBased}%
  \BibitemOpen
  \bibfield  {author} {\bibinfo {author} {\bibfnamefont {I.}~\bibnamefont {Goodfellow}}, \bibinfo {author} {\bibfnamefont {Y.}~\bibnamefont {Bengio}}, \ and\ \bibinfo {author} {\bibfnamefont {A.}~\bibnamefont {Courville}},\ }\enquote {\bibinfo {title} {Gradient‐based learning},}\ in\ \href@noop {} {\emph {\bibinfo {booktitle} {Deep Learning}}}\ (\bibinfo  {publisher} {MIT Press},\ \bibinfo {address} {Cambridge, MA},\ \bibinfo {year} {2016})\ pp.\ \bibinfo {pages} {177--191},\ \bibinfo {note} {\url{http://www.deeplearningbook.org}}\BibitemShut {NoStop}%
\bibitem [{\citenamefont {Bengio}\ \emph {et~al.}(2013)\citenamefont {Bengio}, \citenamefont {Léonard},\ and\ \citenamefont {Courville}}]{bengio2013estimatingpropagatinggradientsstochastic}%
  \BibitemOpen
  \bibfield  {author} {\bibinfo {author} {\bibfnamefont {Y.}~\bibnamefont {Bengio}}, \bibinfo {author} {\bibfnamefont {N.}~\bibnamefont {Léonard}}, \ and\ \bibinfo {author} {\bibfnamefont {A.}~\bibnamefont {Courville}},\ }\href {https://arxiv.org/abs/1308.3432} {\enquote {\bibinfo {title} {Estimating or propagating gradients through stochastic neurons for conditional computation},}\ } (\bibinfo {year} {2013}),\ \Eprint {http://arxiv.org/abs/1308.3432} {arXiv:1308.3432 [cs.LG]} \BibitemShut {NoStop}%
\bibitem [{\citenamefont {Courbariaux}\ \emph {et~al.}(2015)\citenamefont {Courbariaux}, \citenamefont {Bengio},\ and\ \citenamefont {David}}]{NIPS2015_3e15cc11}%
  \BibitemOpen
  \bibfield  {author} {\bibinfo {author} {\bibfnamefont {M.}~\bibnamefont {Courbariaux}}, \bibinfo {author} {\bibfnamefont {Y.}~\bibnamefont {Bengio}}, \ and\ \bibinfo {author} {\bibfnamefont {J.-P.}\ \bibnamefont {David}},\ }in\ \href {https://proceedings.neurips.cc/paper_files/paper/2015/file/3e15cc11f979ed25912dff5b0669f2cd-Paper.pdf} {\emph {\bibinfo {booktitle} {Advances in Neural Information Processing Systems}}},\ Vol.~\bibinfo {volume} {28},\ \bibinfo {editor} {edited by\ \bibinfo {editor} {\bibfnamefont {C.}~\bibnamefont {Cortes}}, \bibinfo {editor} {\bibfnamefont {N.}~\bibnamefont {Lawrence}}, \bibinfo {editor} {\bibfnamefont {D.}~\bibnamefont {Lee}}, \bibinfo {editor} {\bibfnamefont {M.}~\bibnamefont {Sugiyama}}, \ and\ \bibinfo {editor} {\bibfnamefont {R.}~\bibnamefont {Garnett}}}\ (\bibinfo  {publisher} {Curran Associates, Inc.},\ \bibinfo {year} {2015})\BibitemShut {NoStop}%
\bibitem [{\citenamefont {{National Institute of Standards and Technology}}(1990)}]{NIST1990SD19}%
  \BibitemOpen
  \bibfield  {author} {\bibinfo {author} {\bibnamefont {{National Institute of Standards and Technology}}},\ }\href@noop {} {\enquote {\bibinfo {title} {{NIST} special database 19: Handprinted forms and characters database},}\ }\bibinfo {howpublished} {Available at \url{https://www.nist.gov/srd/nist-special-database-19}} (\bibinfo {year} {1990}),\ \bibinfo {note} {accessed: 2025-06-12}\BibitemShut {NoStop}%
\bibitem [{\citenamefont {Blais}\ \emph {et~al.}(2004)\citenamefont {Blais}, \citenamefont {Huang}, \citenamefont {Wallraff}, \citenamefont {Girvin},\ and\ \citenamefont {Schoelkopf}}]{QEDcircuit}%
  \BibitemOpen
  \bibfield  {author} {\bibinfo {author} {\bibfnamefont {A.}~\bibnamefont {Blais}}, \bibinfo {author} {\bibfnamefont {R.-S.}\ \bibnamefont {Huang}}, \bibinfo {author} {\bibfnamefont {A.}~\bibnamefont {Wallraff}}, \bibinfo {author} {\bibfnamefont {S.~M.}\ \bibnamefont {Girvin}}, \ and\ \bibinfo {author} {\bibfnamefont {R.~J.}\ \bibnamefont {Schoelkopf}},\ }\href {\doibase 10.1103/PhysRevA.69.062320} {\bibfield  {journal} {\bibinfo  {journal} {Phys. Rev. A}\ }\textbf {\bibinfo {volume} {69}},\ \bibinfo {pages} {062320} (\bibinfo {year} {2004})}\BibitemShut {NoStop}%
\bibitem [{\citenamefont {Koch}\ \emph {et~al.}(2007)\citenamefont {Koch}, \citenamefont {Yu}, \citenamefont {Gambetta}, \citenamefont {Houck}, \citenamefont {Schuster}, \citenamefont {Majer}, \citenamefont {Blais}, \citenamefont {Devoret}, \citenamefont {Girvin},\ and\ \citenamefont {Schoelkopf}}]{Cooperpair}%
  \BibitemOpen
  \bibfield  {author} {\bibinfo {author} {\bibfnamefont {J.}~\bibnamefont {Koch}}, \bibinfo {author} {\bibfnamefont {T.~M.}\ \bibnamefont {Yu}}, \bibinfo {author} {\bibfnamefont {J.}~\bibnamefont {Gambetta}}, \bibinfo {author} {\bibfnamefont {A.~A.}\ \bibnamefont {Houck}}, \bibinfo {author} {\bibfnamefont {D.~I.}\ \bibnamefont {Schuster}}, \bibinfo {author} {\bibfnamefont {J.}~\bibnamefont {Majer}}, \bibinfo {author} {\bibfnamefont {A.}~\bibnamefont {Blais}}, \bibinfo {author} {\bibfnamefont {M.~H.}\ \bibnamefont {Devoret}}, \bibinfo {author} {\bibfnamefont {S.~M.}\ \bibnamefont {Girvin}}, \ and\ \bibinfo {author} {\bibfnamefont {R.~J.}\ \bibnamefont {Schoelkopf}},\ }\href {\doibase 10.1103/PhysRevA.76.042319} {\bibfield  {journal} {\bibinfo  {journal} {Phys. Rev. A}\ }\textbf {\bibinfo {volume} {76}},\ \bibinfo {pages} {042319} (\bibinfo {year} {2007})}\BibitemShut {NoStop}%
\bibitem [{\citenamefont {Barney}(2025)}]{repo2024}%
  \BibitemOpen
  \bibfield  {author} {\bibinfo {author} {\bibfnamefont {R.}~\bibnamefont {Barney}},\ }\href@noop {} {\enquote {\bibinfo {title} {Natural quantization of neural networks},}\ }\bibinfo {howpublished} {\url{https://github.com/Galitski-theory-group/quantum-nn}} (\bibinfo {year} {2025})\BibitemShut {NoStop}%
\bibitem [{\citenamefont {Lakhdar-Hamina}(2025)}]{repo2025}%
  \BibitemOpen
  \bibfield  {author} {\bibinfo {author} {\bibfnamefont {D.}~\bibnamefont {Lakhdar-Hamina}},\ }\href@noop {} {\enquote {\bibinfo {title} {{Natural-Quantization}: A hybrid classical–quantum neural network framework},}\ }\bibinfo {howpublished} {\url{https://github.com/Galitski-theory-group/natural-quantization}} (\bibinfo {year} {2025})\BibitemShut {NoStop}%
\bibitem [{\citenamefont {Debnath}\ \emph {et~al.}(2016)\citenamefont {Debnath}, \citenamefont {Linke}, \citenamefont {Figgatt}, \citenamefont {Landsman}, \citenamefont {Wright},\ and\ \citenamefont {Monroe}}]{debnath2016demonstration}%
  \BibitemOpen
  \bibfield  {author} {\bibinfo {author} {\bibfnamefont {S.}~\bibnamefont {Debnath}}, \bibinfo {author} {\bibfnamefont {N.~M.}\ \bibnamefont {Linke}}, \bibinfo {author} {\bibfnamefont {C.}~\bibnamefont {Figgatt}}, \bibinfo {author} {\bibfnamefont {K.~A.}\ \bibnamefont {Landsman}}, \bibinfo {author} {\bibfnamefont {K.}~\bibnamefont {Wright}}, \ and\ \bibinfo {author} {\bibfnamefont {C.}~\bibnamefont {Monroe}},\ }\href {\doibase https://doi.org/10.1038/nature18648} {\bibfield  {journal} {\bibinfo  {journal} {Nature}\ }\textbf {\bibinfo {volume} {536}},\ \bibinfo {pages} {63} (\bibinfo {year} {2016})}\BibitemShut {NoStop}%
\bibitem [{\citenamefont {Bishop}(1995)}]{bishop1995noise}%
  \BibitemOpen
  \bibfield  {author} {\bibinfo {author} {\bibfnamefont {C.~M.}\ \bibnamefont {Bishop}},\ }\href {\doibase 10.1162/neco.1995.7.1.108} {\bibfield  {journal} {\bibinfo  {journal} {Neural Computation}\ }\textbf {\bibinfo {volume} {7}},\ \bibinfo {pages} {108} (\bibinfo {year} {1995})}\BibitemShut {NoStop}%
\bibitem [{\citenamefont {Tomesh}\ \emph {et~al.}(2022)\citenamefont {Tomesh}, \citenamefont {Gokhale}, \citenamefont {Omole}, \citenamefont {Ravi}, \citenamefont {Smith}, \citenamefont {Viszlai}, \citenamefont {Wu}, \citenamefont {Hardavellas}, \citenamefont {Martonosi},\ and\ \citenamefont {Chong}}]{benchmark}%
  \BibitemOpen
  \bibfield  {author} {\bibinfo {author} {\bibfnamefont {T.}~\bibnamefont {Tomesh}}, \bibinfo {author} {\bibfnamefont {P.}~\bibnamefont {Gokhale}}, \bibinfo {author} {\bibfnamefont {V.}~\bibnamefont {Omole}}, \bibinfo {author} {\bibfnamefont {G.~S.}\ \bibnamefont {Ravi}}, \bibinfo {author} {\bibfnamefont {K.~N.}\ \bibnamefont {Smith}}, \bibinfo {author} {\bibfnamefont {J.}~\bibnamefont {Viszlai}}, \bibinfo {author} {\bibfnamefont {X.-C.}\ \bibnamefont {Wu}}, \bibinfo {author} {\bibfnamefont {N.}~\bibnamefont {Hardavellas}}, \bibinfo {author} {\bibfnamefont {M.~R.}\ \bibnamefont {Martonosi}}, \ and\ \bibinfo {author} {\bibfnamefont {F.~T.}\ \bibnamefont {Chong}},\ }in\ \href {\doibase 10.1109/HPCA53966.2022.00050} {\emph {\bibinfo {booktitle} {2022 IEEE International Symposium on High-Performance Computer Architecture (HPCA)}}}\ (\bibinfo {year} {2022})\ pp.\ \bibinfo {pages} {587--603}\BibitemShut {NoStop}%
\end{thebibliography}%

\end{document}